\documentclass[10pt,a4paper,dvips,twocolumn,showpacs]{revtex4}

\usepackage[english]{babel}
\usepackage{epsfig}
\usepackage{amsmath, amsfonts, amssymb}
\usepackage{graphicx}

\usepackage{dcolumn}

\usepackage[dvips]{color}

\newcommand{\um}{\overline{u}}

\newcommand{\vm}{v_f}
\newcommand{\Vch}{V_{\chi}}
\newcommand{\Vf}{V_f}

\begin{document}

\title{Self-Sustained Reaction Fronts in Porous Media}

\author{Severine Atis}
\author{Sandeep Saha}
\author{Harold Auradou}
\author{Dominique Salin}
\author{Laurent Talon}
\affiliation{Univ.\ Pierre et Marie Curie, Univ.\ Paris-Sud, CNRS,
Lab.\ FAST, B{\^a}t.\ 502, Campus Univ., Orsay, F--91405, France.}

\today

\begin{abstract}
We analyze experimentally chemical waves propagation in the disordered flow field of a porous medium. The reaction fronts travel at a constant velocity which drastically depends on the mean flow direction and rate. The fronts may propagate either downstream and upstream but, surprisingly, they remain static over a range of flow rate values. Resulting from the competition between the chemical reaction and the disordered flow field, these frozen fronts display a particular sawtooth shape. The frozen regime is likely to be associated with front pinning in low velocity zones, the number of which varies with the ratio of the mean flow and the chemical front velocities.
\end{abstract}

\maketitle

{\it Introduction.}
Interface motion and front propagation are relevant to many processes,
including chemical reactions \cite{scott94}, population dynamics in biology,  flame propagation in combustion \cite{zeldovich38} and chemotaxis \cite{adler66}.
The dynamic of front propagation is well understood in stagnant fluids \cite{fisher37,kolmogorov37}, whereas the effect of fluid flow on front motion is currently of immense interest \cite{abel01,edwards02,pocheau06,schwartz08}.
In this experiment, we consider an autocatalytic reaction between two chemical species which generates a self-sustained reaction front \cite{hanna82,toth97}. These fronts propagate as solitary waves with a constant velocity ($V_{\chi}\propto\sqrt{D_m \alpha}$) and a stationary concentration profile of width ($l_{\chi}=D_m /V_{\chi}$), reflecting both the balance between molecular diffusion $D_m$ and reaction rate $\alpha$.

In heterogenous flow field, non-linear coupling between reaction, diffusion and advection generally induce long-range deformation of the reaction front. For instance, in a parabolic Poiseuille flow \cite{edwards02,leconte03,leconte08}, this coupling leads to a stationary front concentration profile propagating at a constant velocity, which shape and velocity are related to the local flow structure and the boundary conditions.
More complex flows have been addressed in porous media \cite{kaern02,koptyug08} or in cellular flow \cite{schwartz08}. Interestingly, both these experiments have shown the existence of static fronts for flow in direction opposite to the chemical wave propagation.
In this context, we have investigated the interaction of a self-sustained reaction front with the disordered flow field in a porous medium. Our experiments reveal new features of front propagation in porous media due to the use of a wide and transparent quasi-two dimensional design, allowing the flow velocity fluctuations measurements and the direct observation of the reaction fronts. Depending on the flow velocity and direction, the self-sustained fronts can travel downstream, upstream with a constant front velocity but they can also remain static over a range of flow rate values. 

\begin{figure}[htbt]
\begin{center}
\includegraphics[width=8cm]{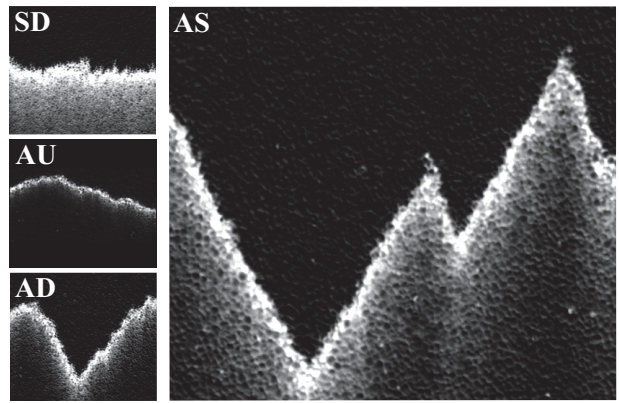}
\caption{\small Typical self-sustained chemical fronts in porous media. The first label corresponds to the mean flow direction, supportive $S$ or adverse $A$ to the chemical wave propagation; the second label denotes the front propagation, downstream $D$, upstream $U$ or static $S$.}
\label{fronts}
\end{center}
\end{figure}

In this Letter, we show visual aspects of frozen reaction fronts in porous media. These static fronts are strongly distorted and display a remarkably straight sawtooth pattern, as shown in Fig. \ref{fronts}; the patterns aspect, the number of peaks and size, evolving with the mean flow rate. We determine first the main characteristics of the flow field such as its correlation length and the spatio-temporal maps of the local velocities along the front as well as its probability distribution function ($PDF$). Analyzing the freezing dynamics then provides a better physical understanding of frozen fronts formation mechanism in porous media.

\begin{figure}[htbt]
\begin{center}
\includegraphics[width=8cm]{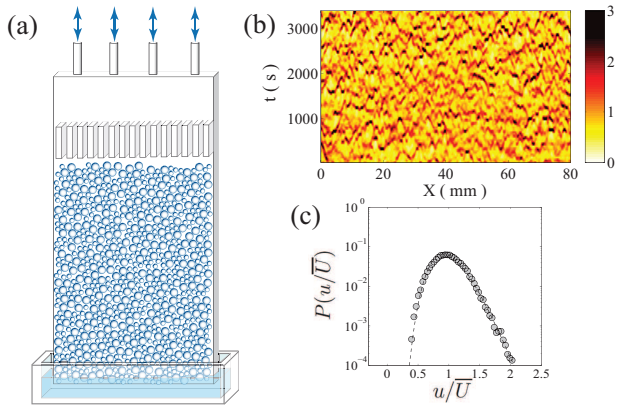}
\caption{(a) Experimental set-up schematic. (b) Spatio-temporal map of the velocity variations measured with dye transport in the porous medium. (c) $PDF$ of the above velocities map (dots) and log-normal fit (dashed line).}
\label{set-up}
\end{center}
\end{figure}

{\it Experimental setup.}
We performed experiments with the Iodate Arsenous Acid (IAA) autocatalytic reaction in iodide (I$^{-}$): $3 \mbox H_{3}\mbox{AsO}_{3} + \mbox{IO}_{3}^- + 5\mbox I^-\longrightarrow 3 \mbox H_{3}\mbox{AsO}_{4} + 6 \mbox I^-$. For the concentration values used here ($[\mbox{IO}_{3}^{-}]_{0}=7.5$ mM, $[\mbox H_{3}\mbox{AsO}_{3}]_{0}=25$ mM), the arsenous is in excess \cite{hanna82}. The front position is localized by the transient generation of iodine during the reaction and marked by polyvinyl alcohol (see Fig \ref{fronts} for instance) \cite{boumalham10}.

The porous medium consists of $50\%$ mixture of $1.5$ and $2 \; mm$ glass spheres (porosity: $48 \pm 2 \%$). They are packed in a transparent rectangular cell ($30 \times 10 \times 1 \,cm^{3}$) as shown in Fig. \ref{set-up}.
To overcome buoyancy effect \cite{pojman91}, we increase the viscosity to $10 \; mPa.s$.
The cell is filled with reactant solution and the chemical reaction is initiated at its bottom end. In the absence of flow, the reaction develops into a flat horizontal front propagating upward at the chemical velocity $V_{\chi}\simeq 10\, \mu m/s$ and $l_{\chi}\simeq 100 \, \mu m $. When the front reaches the desired vertical position, the flow is switched on and the fluid can be either sucked out or injected parallel to the vertical from the top of the cell. 
We extract the location $h(x)$ of the chemical pseudo-interface at a given transverse distance $x$ and define the front velocity by $v(x,t)=(h(x,t+\Delta t)-h(x,t))/\Delta t$; $\Delta t$ is such that the variation of $h(x)$ is larger than the correlation length.

The net flow is characterized by its average value, $\overline{U}$ and the characteristic length of the velocity fluctuation $l_d$, is inferred from tracer dispersion experiments \cite{hulin88,bacri87}: $l_d=(1.8 \pm 0.1) \; mm$ which is close to the mean particle diameter, in agreement with  \cite{koch85}.
\begin{figure}[htbt]
\begin{center}
\includegraphics[width=7.5cm]{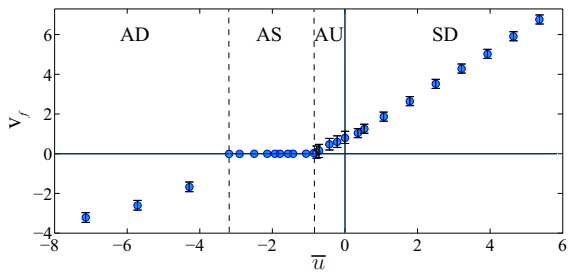}
\caption{\small Front velocity $v_{f}$ versus mean flow velocity in the porous medium, $\um$.}
\label{vf}
\end{center}
\end{figure}
Tracer front velocity fluctuations $u(x,t)$ are shown in Fig. \ref{set-up}. The spatiotemporal map displays the flow heterogeneities of characteristic size larger than $l_d$ and the flow velocities $PDF$ is asymmetric and well fitted by a log-normal distribution (solid line).

\begin{figure*}[htbt]
\begin{center}
\includegraphics[width=17.5cm]{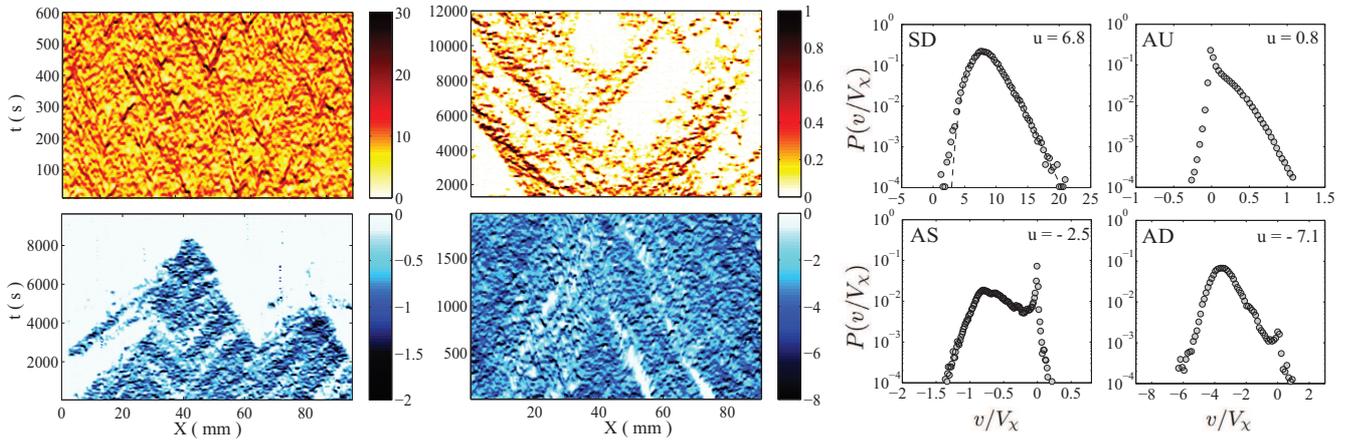}
\caption{Spatiotemporal representation of $v(x,t)$ for the four regimes. Color code intensity represents the magnitude of the velocity; red: $v>0$, blue: $v<0$. White regions corresponds to $v=0$. The corresponding $PDF$s of $v(x,t)$ (dots) and log-normal fit (dashed line).}
\label{pdfDSF}
\end{center}
\end{figure*}

{\it Front velocity.}
We performed experiments over a wide range of mean flow velocities and measured the reaction front velocity. The fronts propagate with a velocity slightly fluctuating around a mean value, noted $V_f$.
Counting as positive velocities in the direction of the reaction front propagation, $\overline{U}>0$ corresponds to supportive flow ($S$) and  $\overline{U}<0$ to adverse flow ($A$). The mean front and mean flow velocities are, now, normalized by $\Vch$; $\um=\overline{U}/V_{\chi}$ and $\vm=\Vf/V_{\chi}$. As can be seen from Fig. \ref{vf}, $\vm$ depends on both the value and direction of $\um$ and one can identify four different regimes of front propagation; the corresponding front's aspects are shown in Fig. \ref{fronts}.
For supportive flow ($\um>0$), the front propagates downstream, i.e. in the flow direction ($\vm>0$). In this supportive downstream regime ($SD$) , $\vm$ varies almost linearly with $\um$  with a slope larger than one. As demonstrated in Poiseuille flow \cite{edwards02,leconte03}, this may be attributed to the selection of the fastest streamline by the reaction front when propagating downstream.
For adverse flow ($\um<0$), the competition between the chemical wave and the flow field leads to more complex behaviors. 
For a small adverse flow rate  $-1\lesssim \um \leq 0$, the chemical wave overcomes the flow leading to upstream propagating fronts ($AU$, $\vm>0$); $\vm$ varies smoothly across the regimes $SD$ and $AU$ and approaches zero. 
The most salient regime is observed below $\um \simeq -1$. The reaction fronts are static without any fluctuations and display frozen sawtooth like patterns (Fig. \ref{fronts}). They are observed over a range of mean flow velocities $\um \in [-3.2,-1]$ which corresponds to the plateau domain ($\vm = 0$, $AS$ regime).
Finally, for $ \um \lesssim -3.2$, the flow is strong enough to overcome the chemical wave and so that the front moves downstream, opposite to the chemical wave velocity ($\vm<0, AD$). Its velocity varies again almost linearly with $\um$. However, in contrast to the $SD$ regime, the resulting slope is smaller than one, which is in agreement with similar experiments using Hele-Shaw cells \cite{edwards02}.\\
Static fronts have been observed in Poiseuille flow in Hele-Shaw cells \cite{edwards02,leconte03}, but they were obtained only for a unique value of $\um$. In contrast, the velocity diagram determined here is similar to that observed in other contexts. For instance, the domains of observation of frozen fronts has been determined in combined cellular and mean opposite flow \cite{schwartz08} as a function of the vortex and the mean adverse flow intensities. Static fronts in packed-bed reactors with different chemistries\cite{kaern02,koptyug08} have also been reported for adverse flow. Quite remarkably the reaction fronts exhibit similar behaviors despite the important differences between these systems. Furthermore, one common feature of all these experiments is the heterogeneity of the flow and shows here the key part the flow spatial structures play in frozen fronts formation. 

{\it Spatio-temporal dynamics.}
One of the major motivations of these experiments, is to understand how the flow velocity fluctuations alter the front propagation and account for the different regimes. This can be inferred from the fronts dynamics; Fig. \ref{pdfDSF} displays the spatio-temporal fluctuations of $v(x,t)$ and the associated $PDF$s in the different regimes.
For $SD$ fronts, the mean flow and the chemical wave are in the same direction, therefore the measured local front velocities, $v(x,t)$, have positive values during the experiment. The $PDF$ of these velocities has a log-normal shape with almost the same standard deviation as the one measured previously for the hydrodynamic flow field. This suggests that the hydrodynamic fluctuations dominate the front dynamic in this regime.
The reaction front propagation is strikingly different in adverse flow regimes. The $PDF$s are modified indicating a more complex dynamical response than just the addition of the chemical velocity to the flow velocity distribution. The spatio-temporal maps exhibit zero velocity regions. This is a signature of transient static portions of the fronts, resulting in a local maximum at $v=0$  on the $PDF$s.
In the $AU$ regime, one can identify long range lateral correlation of the velocities as dark regions on the spatio-temporal. They indicate transversal propagation of the reaction at constant rate. At a given location, when the front stops, it remains static until laterally traveling parts of the front reach these waiting regions. In this configuration, the fronts propagate upstream, however the $PDF$ displays a few negative velocities. Those values can be attributed to the locations where the flow velocity is locally larger than the chemical wave velocity. 
In the $AS$ regime, one can note that the final static pattern is achieved after a transient phase. Therefore, the spatio-temporal map describes the freezing dynamics of the front. Velocity fluctuations are negative, illustrating the fact that the front is mostly pushed back by the flow until it pins locally to a point-like region (at $X = 63\, mm$ in Fig. \ref{pdfDSF}). The final front structure displays a sawtooth shape (Fig. \ref{fronts}), where each pinning point is at the vertex of a peak. However, before achieving the final pattern, transitorily pinned parts of the front are visible on the map as white stripes. In contrast to the $AU$ regime, this demonstrates the long transversal correlation of zero velocity regions over time. Hence, the front develops transient static parts during the whole experiment until it reaches an entirely frozen state. 
Finally, in the $AD$ regime, fronts move downstream all along the experiments, and thus in this regime the flow seems to dominate the propagation of reaction. Yet a small peak at $v=0$ on the $PDF$ indicates the persistence of transitorily pinned parts of the fronts, which progressively vanish as the mean adverse flow velocity is increased.

\begin{figure}[htbt]
\begin{center}
\includegraphics[width=8.5cm]{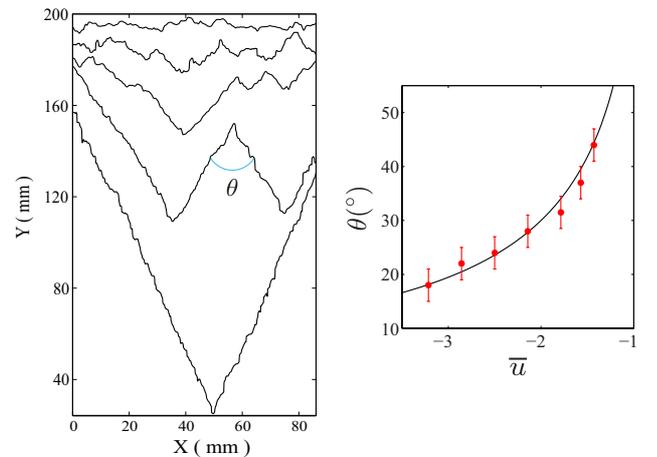}
\caption{Left: Frozen fronts aspects for different mean flow rate: from top to bottom, $\um=-1.4,-1.6,-1.9,-2.5,-3.2$. Right: Measurement of $\theta$ versus $\um$; solid line, theory.}
\label{steady_fronts}
\end{center}
\end{figure}

{\it Discussion.}
The dynamical study has revealed pinning regions of the reaction fronts in adverse flows. They play a determining role in the freezing mechanism for the plateau regime ($AS$). Experiments on stationary Belousov-Zhabotinsky pulses have been reported in packed bead reactors \cite{kaern02} and linked to an "excited stagnant pockets" mechanism. When the reaction front reaches these regions, they act as point source of chemical reaction. Producing and spreading catalyst, these pockets can sustain the reaction front in their vicinity. Although the chemical reaction is very different in that case, one can follow a similar argument in our experiments such that these pinning zones may correspond to stagnant flow pockets.

However this argument cannot account for reaction fronts moving downstream for sufficiently large adverse flow. This indicates that pinning points must be associated with weak but generally non-zero velocity zones, i.e. $-U(\vec{r})\lesssim\Vch$. Depinning of the front then occurs when the local flow velocity reaches a critical value which pushes the reaction out of this region. In this case, increasing the mean velocity results in reducing the number of low velocity zones required for pinning.
This mechanism is supported by visual observations displayed on Fig. \ref{steady_fronts}. Several frozen patterns recorded for different $\um$ values in the plateau domain are plotted and one can identify the location of the pinning points with the vertex of the front's peaks. Note that the two lateral boundaries can also be considered as pinning zones.
As the flow velocity increases, the number of pinning points decreases until they vanish completely (except the two cell lateral boundaries) for $\um=-3.2$. 
This suggests that, statistically, the chemical reaction finds fewer pockets which can remain excited when $\um$ increases. 
The shapes of the final static fronts depend on the distribution of these pinning points but also on the initial location of the front.

Away from these zones, the flow velocity is much higher than $\Vch$, and yet, the fronts reach a frozen state all over the porous medium.
In order to understand these frozen fronts, one has to deal with the two flow regimes already analyzed in simple flows \cite{edwards02,leconte03}.
In the so called mixing regime, the size of the flow extension ($l_d$) is much smaller than the chemical front width, $l_{\chi}$, molecular diffusion has enough time to transversally mix the reactants leading to the Galilean invariance: $V_{f}=\bar{U}+V_{\chi}$.
Therefore, static fronts are possible for a unique flow rate $\bar{U}=-V_{\chi}$; even with the flow distribution of Fig. \ref{set-up}, it is not enough to account for the plateau. In the opposite, ($l_d >>l_{\chi}$), thin front eikonal limit \cite{williams85}, at each point of the front surface, the normal component of the local interface velocity satisfies :
\begin{equation} \label{eikonal}
\vec{V_{f}}(\vec{r})\cdot\vec{n}=  V_{\chi} + \vec{U}(\vec{r})\cdot\vec{n}+D_m \kappa
\end{equation}
where $\vec{U}(\vec{r})$ is the local fluid velocity at the front position $\vec{r}$, $\vec{n}$ is the local unit vector normal to the interface and $\kappa$, the curvature of the interface. In this regime, frozen fronts can adjust their local inclination and curvature to accommodate larger local flow velocities and satisfy: $V_f=0$ in Eq. \ref{eikonal}.
Hence, increasing the flow intensity from $\um=-1$ to $\um=-3.2$ results in larger adverse flow leading to more inclination and bending of the front around pinning zones. Note that, we are here in an intermediate regime ($l_d \sim 10 l_{\chi}$) where reactants concentrations variations occur over a few $l_{\chi}$, however the thin front limit still explains the aspect of the frozen states. 
In Fig. \ref{steady_fronts}, the teeth angle $\theta$ decreases with $|\um|$. Its value is in agreement with: $\Vch+\overline{U}\sin(\theta/2) = 0$ (solid line), which derives from Eq. \ref{eikonal} with the curvature term neglected on the straight parts of the front.
Particularly, this pattern is analogous to the V-shape displayed on flame fronts, strongly inclined in order to counterbalance high flow rate \cite{echekki90}.

Finally, we have shown that the transient static portions of the fronts observed on the spatio-temporal maps, correspond to transitory pinning zones. During the experiments, the reaction fronts are temporally pinned at these locations.  They form local static parts without leading to a completely frozen pattern and depin suddenly causing burst-like events. The irregular propagation of the fronts in $AS$ and $AD$ regimes is hence due to successive rearrangements of the fronts when they travel from one low velocity zone to another one. In these specific regimes, the fronts propagate through local pinning and depinning events whose statistical properties \cite{santucci11} deserve further investigations aided by numerical simulations.

{\it Conclusion.}
We have experimentally investigated the propagation of chemical waves coupled to the disordered flow of a model porous medium. Depending on the mean flow direction and intensity, the reaction fronts exhibit four different regimes. Our experiments show the key role played by the flow heterogeneities on the chemical front dynamics. An interesting freezing regime is observed and explained by front pinning to low flow velocity regions. More specifically, we have analyzed the complex dynamic features involved in this pinning process. We thus expect that the domain of existence of this regime strongly depends on the flow heterogeneities distribution and further numerical simulations should be performed to analyze this dependence.

{\it Acknowledgements.-}
We thank J.-P. Hulin for stimulating discussions. The research was partly supported by CNES, RTRA ''Triangle de la physique''. S.S  was supported by the Initial Training Network (ITN) ''Multiflow''.

\end{document}